# An Analysis of the 'Blind Variation and Selective Retention' Theory of Creativity

Liane Gabora
University of British Columbia

For correspondence regarding the manuscript:

Liane Gabora
Department of Psychology
University of British Columbia
Okanagan Campus, 3333 University Way
Kelowna BC, Canada V1V 1V7
Email: liane.gabora@ubc.ca
Tel: 250-807-9849
Fax: 250-470-6001

ABSTRACT

Picasso's Guernica sketches continue to provide a fruitful testing ground for examining and assessing the Blind Variation Selective Retention (BVSR) theory of creativity. Nonmonotonicity—e.g. as indicated by a lack of similarity of successive sketches—is not evidence of a selectionist process; Darwin's theory explains adaptive change, not nonmonotonicity. Although the notion of blindness originally implied randomness, it now encompasses phenomena that bias idea generation, e.g. the influence of remote associations on sketch ideas. However, for a selectionist framework is to be applicable, such biases must be negligible, otherwise evolutionary change is attributed to those biases, *not* to selection. The notion of 'variants' should not be applied to creativity; without a mechanism of inheritance, there is no basis upon which to delineate, for example, which sketch ideas are or are not variants of a given sketch idea. The notion of selective retention is also problematic. Selection provides an explanation when acquired change is not transmitted; it cannot apply to Picasso's painting (or other creative acts) because his ideas acquired modifications as he thought them through that were incorporated into paintings and viewed by others. The generation of one sketch affects the criteria by which the next is judged, so sequentially generated sketches cannot be treated as members of a generation, and selected amongst. Although BVSR is inappropriate as a theoretical framework for creativity, exploring to what extent selectionism explains the generation of not just biological form but masterpieces such as Picasso's Guernica is useful for gaining insight into creativity.

There has been ongoing discussion in this journal of the theory, originally proposed by Campbell (1960), that creative ideation is a two-stage mental process involving 'blind' variation and selective retention, abbreviated BVSR. The idea is that the creator mutates a given idea a multitude of different ways, selects the fittest variant(s), and repeats this procedure until an adequately creative idea results. For the last few decades, BVSR—sometimes referred to as the Darwinian view of creativity—has been most closely associated with Dean Simonton, who has developed it in interesting ways to incorporate findings from the literature, and amassed considerable data in support of it (1995, 1998, 1999a,b, 2004, 2005, 2007a, in press). Darwin's theory of natural selection is the canonical example of a *selectist* theory, according to which change in a particular domain proceeds in two stages, the first of which entails the chance or unstrategic generation of variants, the second of which entails selection amongst them.

In a previous paper in this journal, Simonton (2007a) summarized the theory and provided putative support for it based on an analysis of the 45 numbered and dated sketches produced by Picasso in the making of his well-known painting, Guernica (Simonton, 2007a). Commentators had the opportunity to respond to that paper (Beghetto & Plucker, 2007; Gabora, 2007; Mumford & Antes, 2007; Weisberg & Hass, 2007), followed by replies from Simonton (Simonton, 2007b). This paper addresses concerns with the theory that remain. We begin with issues pertaining to the 'blind variation' phase of the creative process, followed by issues pertaining to the 'selective retention' phase. We then discuss issues surrounding the Darwinian heritage of BVSR. Finally, we assesses the case for BVSR as a general theory of creativity.

## Blind Variation

### The Issue of Randomness Revisited

Simonton uses the term 'blindness' to mean (1) that variants are generated by chance, random or 'pseudo-random'[1] processes typically occurring below the threshold of awareness, and (2) the creator has no subjective certainty about whether an idea is a step in the direction of the final creative product. He has loosened up on his claim that creative novelty is generated randomly; it is now conceded that expertise, remote associates, and strategy play substantial though subsidiary roles in the generation of creative ideas (Simonton, 1995, 1999a,b, 2003, 2004, 2007a, 2010). An example is allowance for the influence of remote associates. This move puts BVSR into stronger alignment with what we know about how the creative process works. The problem with it, however, is that not only are logic and remote associates not *part* of a selectionist theory, they are *inconsistent* with the basic tenets of Darwinism. Darwin's theory of natural selection explains change over generations in the distribution of heritable variation due to differential selection. It explains why black moths become more abundant on black trees that camouflage them from predators. Natural selection *assumes* randomly generated variation; that is, if Darwin's theory is to be applicable to a phenomenon, it is required that agents of change at the generation phase, i.e. *prior to* selection, be *negligible*. Deviations from randomly generated variation render selectionism inapplicable as an explanatory framework because evolutionary change in the distribution of variants over time is then attributed to the *nature* of those biases, *not* to differential selection amongst variants. A generation process that is biased by strategy, learning, remote associates, and so forth, is *not* effectively simulated by a random combinatorial model.

Simonton (2007b) writes: "No Darwinian theory of any kind… would ever assert that Darwinism requires random selection! Certainly this is not a premise of the Darwinian theory of creativity" (p. 391). But it has never been claimed that Darwinism requires random *selection*. It requires random *variation*; that is standard Darwinian theory. Biological variation is not genuinely random—for example, we can trace the source of some mutations to various causal agents—but the assumption of randomness generally holds well enough to serve as a useful approximation.[2] More precisely, it *is* possible for the theory of natural selection to be applicable even if the underlying process is not random. However, in that case, although not genuinely random, the process must be *approximated by* a random distribution. To the extent that expertise, remote associates, and so forth, influence creative thought, variation is not generated randomly but biased. Thus change in the creator's ideas over time is attributed to those cognitive biases themselves, *not* to selection amongst randomly generated variant ideas.

Simonton (2007a) contrasts those who view creativity as a systematic, straightforward, logical process of heuristic search (Hayes, 1989; Simon, 1973, 1986; Weisberg, 1992) with those who view it as chaotic, unpredictable, and even inefficient (Eysenck, 1995; Simonton, 1999a), claiming that a Darwinian theory of creativity supports the second view. However, natural selection *is* a kind of heuristic search, specifically a distributed hill-climbing technique. It operates by generating variants through a process that can be approximated by a random distribution, such that any *given* variant is unlikely to be advantageous, but offsets that by generating *lots* of them, such that at least one of them is bound to be fitter than what came before.[3] Genetic algorithms—computer programs that solve problems by embodying the logic of natural selection—are systematic, straightforward, and logical. Thus it is not clear that a Darwinian view of creativity supports the second view of creativity over the first.

Moreover, with respect to creativity, it is not necessary to posit numerous variant solutions generated through chance processes. Because of how human memory encodes information, there is no need for large numbers of possibilities generated by chance processes (Gabora, 2010). In a sparse, distributed, content-addressable memory, items that share features can potentially access one another even if their relationship has never been explicitly noted. The associations one makes reflect one's knowledge and experiences and how they are organized, as well as the present context. Creative minds work not by generating many possibilities blindly but by generating *few* possibilities *intelligently,* such that they are more likely than chance to be adaptive.

**Nonmonotonicity**

Simonton claims that, as a consequence of blindness, the creative process exhibits nonmonotonicity—deviations from gradual improvement—as evidenced by false starts, wild experiments, and backtracking (Simonton, 1999b; cf. Weisberg, 2004). A source of putative support for this is Picasso's Guernica sketches. Simonton (1999b) shows that the sequence of sketches exhibits backtracking: experiments that were not developed further, and did not end up in the final painting. (For example, one unused sketch for the part of the painting containing a mother with a dead child has her carrying the child up a ladder.) Simonton (2007a) isolated each figure from each sketch for a total of 79 stimuli, and asked participants to put them into the order that best demonstrates progress toward the finished product (*i.e.,* the order in which they were most likely made). The participants showed strong agreement in their ordering (composite progress score = .85), and concurred that this ordering was nonmonotonic.

One could object here that the Guernica sketches are not characteristic of a typical creative endeavor; successive drafts of a song or novel or machine for instance probably exhibit more obvious convergence toward a final product than the Guernica sketches (Gabora, 2007). One could also object that, for many such problems, improvement with respect to one criterion may decrease performance with respect to other criteria, such that the best one can do is find pareto optima, and whether the process looks nonmonotonic or not depends on which criteria are taken into account. Yet another objection is that in this kind of analysis, instances of merely switching from a sketch for one part of the painting to a sketch for another part falsely appear as instances of backtracking (Gabora, 2007; Weisberg & Hass, 2007), as will switching from one component of a sketch to another. Moreover, there are problems arising from the kinds of similarity judgments Simonton uses as the basis for determining instances of backtracking. These kinds of similarity judgments have long been known in psychology to be problematic (Goldmeier, 1972; Tversky, 1977; see Medin, 1989, for a critique of Tversky). Similar with respect to *what?* When there are multiple features or dimensions involved, with multiple levels of hierarchical relationship amongst them, as is generally the case with a creative work, there are many ways of weighting these features and their interrelationships to arrive at a similarity judgment. It is doubtful that other people's judgments about Picasso's work would match Picasso's own.

One might also object that a lack of continuity in the sketches need not reflect a lack of continuity in the underlying thought process, for undoubtedly as Picasso made externally visible progress toward completion of one part of the painting, he was subconsciously busy with another part (Gabora, 2007). Nonmonotonicity of sketches reflects that a sketch itself must not only be a satisfying realization of one's idea for that element of the painting, but it must also contribute to the painting as a whole in a satisfying way. Thus one moves on to a new sketch that is different from the previous one not because one's conception of the painting as a whole changes in a discontinuous manner, but because the sketch only actualizes a fragment of the whole (for a similar argument see Doyle, 2008). In my commentary I liken what the creator is aware of to the tip of an iceberg, and Simonton responds by saying that all you need to see is the tip of the iceberg to determine what direction it is going. He does not attempt to show, and indeed it is not obvious, how his elaboration of the metaphor supports his claims. Detecting what direction the tip of an iceberg is drifting tells you something about the overall direction of the rest of the iceberg (though not everything, for calm movement at the surface may obscure the wrenching apart of different pieces of the iceberg taking place below.) But it tells you nothing about its shape, or internal structure, or how that shape and internal structure are changing with time. What appears to be a small, flat iceberg may actually be large, jagged one. Similarly, what appears superficially to be backtracking at the behavioral level may be necessary to moving forward at the cognitive level.

Let us see how this could work using the Guernica example. The series of thoughts or ideas that Picasso had as he worked on Guernica can be referred to as $i_{t1}, i_{t2}, i_{t3}, \ldots i_{t100}$. Some of these—let us say this group includes $i_{t10}$, $i_{t15}$, and $i_{t100}$—manifested as sketches. We refer to these outputs as $o_{t10}$, $o_{t15}$, and $o_{t100}$, with $o_{t100}$ being the composite sketch he finally settled on and from which Guernica was painted. These thoughts were likely half-baked or ill-defined, existing in what have been referred to as *states of potentiality* (Gabora, 2005), and through the process of actualizing them as sketches their details became more well-defined. One can view each sketch as an experiment that forces the conception of one portion of the painting to shift from potential to actual, *i.e.*, it takes a form that is concrete (albeit uncommitted in the sense that it will not necessarily be part of the final painting). For example, the process of having thought $i_{t10}$ and manifesting it as sketch $o_{10}$ may have enabled Picasso to get a better sense of what effect it would have to portray the mother carrying a dead child. This shifted the state of Picasso's internal model of the world, or *worldview,* which in turn modified his conception of how the painting should look and how to achieve it, giving rise to thought $i_{t11}$. Thought $i_{t11}$ shifted the state of his worldview, resulting in thoughts $i_{t12}$, $i_{t13}$, $i_{t14}$, and $i_{t15}$, which gave rise to sketch $o_{t15}$. And so forth. It may often be the case, but it is not automatically the case, that a final sketch is more similar to a late sketch than an early one. Let us say that $o_{t15}$ is the unused sketch of the mother climbing a ladder with her dead baby. If a discontinuity is observed in the series of observable outputs—as it is here because $o_{t100}$ is less similar to $o_{t15}$ than to $o_{t10}$—this simply means that thought $i_{t10}$ preceded thought $i_{t15}$ in the sequence of thoughts that culminated in the worldview being in a state in which it could generate thought $i_{t100}$ resulting in sketch $o_{t100}$.

There are, however, even more serious problems with these results and how they have been interpreted. Simonton claims that they support the Darwin-inspired thesis that ideas are generated through trial and error, in large numbers, some of which are selected and many of which are not, and that one variant can be quite different from another. However, it is not intra-generational variation, but change from one generation to the next that natural selection provides an explanation for, and as mentioned previously, BVSR has not been developed to the extent that it incorporates a means by which traits are passed on from one generation to the next, *i.e.,* a mechanism of inheritance.

Even if some workaround is found that avoids all these problems, the most serious problem of all with this study is that a Darwinian theory does not even predict that the data should be nonmonotonic. Indeed the biological evolution process that inspired BVSR has in place numerous mechanisms that keep nonmonotonic change to a *minimum*. This includes lack of inheritance of acquired characteristics (*e.g.,* you don't inherit your mother's tattoo), as well as proofreading enzymes, spontaneous abortion, and speciation (*e.g.,* when organisms become too different they can no longer mate). These multiple, multi-level mechanisms for *avoiding* sudden extreme change are *why* there is an equilibrium there to punctuate. (It is precisely the extraordinary resistance of biological lineages to change that necessitated Darwin's theory.) Thus experimental evidence of nonmonotonicity does not support a Darwinian theory of creativity. Note that Darwin's explanatory framework is silent as to whether variation *at the individual level* is monotonic or not. That is not what it is a theory of; it is a theory of the evolution of species.[4]

Similar problems arise with using evidence of nonmonotonicity in lifespan creativity as support for BVSR. A fundamental problem with these studies is the assumption that the finding of no significant improvement in the quality or quantity of ideas across the lifespan supports a Darwinian theory of creativity. Darwin's theory of natural selection was put forth to explain why organisms become more adapted to their environment over time, so why would a *lack* of improvement *support* a Darwinian theory? What in Darwinian theory the creator's age maps on to is not explained. Does age, for example, map on to the number of generations? Indeed almost no attempt is made to meaningfully map the basic concepts of natural selection (e.g. speciation, inheritance) on to creativity.

**Applying the Concept of Variation to Creativity**

Simonton (2007b) writes: "Gabora seems to confound variation with change. The former concerns single variants emitted by an individual, whereas the latter concerns the properties of the population of individuals" (p. 390). I am not confounding variation with change; the term change can be applied at either the individual level or the population level. However, this sentence brings to light several other problems with BVSR. First, there are problems with applying the term 'variants' to creativity. Organisms are variants if they are descendents of a common ancestor. Even in Darwin's day it was known that if two organisms had a common mother, or grandmother, they were more likely to share traits. It is now possible to objectively measure what percentage of the genomes of two species or two organisms overlap, and make conclusions about their degree of genetic relatedness. But with respect to the creative process, wherein any idea can combine or 'mate with' any other, how do you draw the line between what is and what is not a variant of a given idea? In a world where thoughts of cars may directly or indirectly play a role in the creation of stop signs, toy trucks, paintings of stop signs, songs about toy trucks, and so forth, on what basis can it be said that any one idea is not a variant or descendent of any other? Is the article I read by you a year ago an ancestor of this one? As Vetsigian, Goldenfeld, and Woese (2006) put it, when ancestry arises through acquired change as opposed to inherited change, descent with variation is not genealogically traceable because change is not delimited to specific ancestors but affects the community as a whole.

In a recent article, Simonton (2010) claims that there need be only two variants for BVSR. Not only is the reader not provided with any means of determining what does and does not count as a variant of a particular idea, but it is not clear whether "two variants" means that there are many ideas with trait *a* and many ideas with trait *b*, or only two ideas in total, one with trait *a* and one with trait *b.* Generally the phrase "two variants" would be used to refer to the first, but Simonton's writing implies the second. If there are really only two ideas in total, then clearly evolution through a selectionist process—in which the smarts is in the *selection* rather than the *generation*, and evolution is a function of change is in the distribution of variants across generations—is going to be *incredibly* inefficient! (As mentioned earlier, a selectionist process is only efficient with large numbers of entities because then there is a reasonable hope that at least one will be effective.)

Similarly confusion arises with application of the term 'population' to creativity. When Simonton (2007a) writes that the term 'change' "concerns the properties of the population of individuals" (p. 390), he implies that the term 'population' should be applied to the set of possible ideas, or actual ideas, in an individual's mind. A population is a set of interbreeding organisms. The term is useful when there are constraints on which entities can combine or interbreed. With respect to a mind, since as we have seen there are no constraints on what ideas can combine or 'interbreed', the concept does not apply.

## Selective Retention and Elaboration

### Sequentiality and Changing Selection Criteria

Simonton rightly states that in creative tasks there is often "no fixed goal". The goal may, at least initially, be ill-defined, and there may be no clear conception of how to achieve it. This means that the selection criteria are subject to modification. In a selectionist process, it is assumed that members of a generation are exposed to the same selection criteria. To the extent that selection criteria change between sequentially generated variants, the variants cannot be considered part of the same generation, i.e. they are not *selected amongst*. Of course, selection criteria are *always* changing; thus to some extent the assumption is even too strict to give a perfectly accurate model of biological change. However, the assumption is *particularly* dangerous if the existence of the first variant changes the selection criteria that affect the survival and reproduction of the second. With respect to creativity, as Simonton (2007b) concedes, in creative thought variants *are* generated sequentially:

> "Unlike what occurs in biological evolution and genetic algorithms, where variants in a given breeding season or generational cycle can be emitted simultaneously, the creative individual almost invariably has to generate the variants sequentially"… (p. 389-390).

In fact, in creative thought, each successive idea alters the context against which the next is assessed, if only because having tried one thing you know it is not right so you try something else (Gabora, 2005). Therefore, never are two evaluated with respect to the same criteria, and thus there is no consistent basis for selecting amongst them, as is required for natural selection.

Simonton (2007a) attempts to get around this, claiming, "When Picasso compiles a series of sketches with respect to a given figure, a later sketch in the series does not necessarily replace or supplant an earlier sketch in the series. On the contrary, the more recent sketch is often merely added to the earlier sketches, thereby increasing the number of ideational variants" (p. 389). He continues:

> They are sequential in the sense that the variations have to be generated in a particular order. But they are simultaneous in the sense that the production of variation n_1 does not automatically mean that variation n has been selected out of the running. The earlier variation may still survive as a potential solution to the problem. … So the product of the sequential variation is frequently the accumulation of two or more simultaneous variants that can then undergo selection. (p. 390)

However, that is not the sense that could make a variation-and-selection framework applicable, and the extent to which a new sketch was *viewed* by Picasso (or Simonton) as replacing a previous one or adding to the pile of sketches is irrelevant. What matters is whether sketch *n* alters the conception of the task and where to go next, i.e. influences the selection criteria (what in genetic algorithms is referred to as the fitness function) by which sketch *n+1* is judged. It's the fact that in order to have generated *n+1* you had to first have generated and evaluated *n* that renders it impossible to consider them members of the same generation. Even if Picasso viewed it as adding to the pile of sketches, during the act of sketching it he had thoughts about each figure, and how it relates to other figures, and the implications of these patterns of interrelationship for other potentially distant parts of the painting. Thus his conception of both the sketch, and the painting as a whole, including what he is trying to achieve, and how to achieve it, change over the course of making the sketch. After this sketch, the criteria by which any subsequent sketch is evaluated are not the same as they were before.

Simonton (2007a) then addresses a related concern brought up in my commentary (Gabora, 2007):

> [H]er objection concerns the fact that memories are never retrieved exactly, but rather they are reconstructed. … Creators do not always have to rely on memories. They have supremely accurate records of their previous thoughts. Picasso, Leonardo, and other artists had their sketches and sketchbooks, Darwin, Faraday, Beethoven, and many other creators had their notes and notebooks. Is Gabora willing to say that creativity is Darwinian in at least these occasions?" (p. 391).

The answer is no. External artifacts are, as many have noted (e.g. Clark & Chalmers, 1998; Donald, 1991), still a form of memory, albeit external rather than internal. Upon returning to them they are re-experienced in a new context that is affected in a multitude of subtle ways by the current situation and all the intervening events since the idea was first recorded, which may throw a new light on it or put it in a new perspective. That is exactly the reason they may lead to insight upon the return to them though they did not when they were first recorded. Consider the invention of the beanbag chair. Faced with the task of inventing a chair for kicking back, the designer may feel stumped. After going home and throwing beanbags to his baby, he comes back to work the next day and takes another look at the design he had been working on the day before. It's the same design as yesterday, but after throwing beanbags he conceives of it differently. He realizes that if small beanbags conform to the shape of one's hand, a big beanbag could conform to the shape of one's whole body. It isn't the sketches or designs that make up a creative *thought* process, but the creator's conceptions of them, and these conceptions are affected not only by external memory aids but by a variety of internal and external factors.

Simonton (2007a) continues:

> The second problem is that under most circumstances the reconstructed memory is a reasonable approximation to the original idea. It may be distorted in a number of ways, but under most conditions it will be recognizably similar to what was initially stored. (p. 391)

What matters is that it is different *at all,* because it is exactly these differences that make it a new thought, and that give it the opportunity to have an impact that it did not have when it was first recorded. Simonton then asks, "Is Gabora saying that any departure whatsoever from the original idea automatically invalidates the Darwinian nature of creativity?" (p. 391). Technically, what invalidates the Darwinian framework is if generation of the first one in any way changed the criteria by which the second one is evaluated, or if the first one had to be generated in order for the second one to have the form it has.

**Transmission of Acquired Traits**

Simonton explicitly incorporates the *elaboration* of ideational variants as part of BVSR. For example it is assumed, reasonably enough, that as Picasso thought about his painting of Guernica, his ideas about it changed, and these modifications made their way into the painting, and eventually were witnessed and appreciated by viewers of the painting. Unfortunately, this is highly problematic for a selectionist view of creativity. Natural selection is inapplicable to processes in which there is *non-negligible transmission of acquired characteristics* because *intra*-generational change due to the accumulation of acquired traits overwhelms the slower *inter*-generational mechanism of change due to selection (Gabora, 2005). The transmission of acquired change is rampant with respect to creative ideas. We are constantly transforming ideas in ways that reflect our internal models of the world, altering or combining them to suit our needs, perspectives, or aesthetic sensibilities, and transmitting these modified versions (i.e. versions with acquired traits) to others. The exchange of modified 'lineages of thought' is ubiquitous in human society. If creative ideas *did* evolve through natural selection, then if you heard an idea from a friend, and you reframed it and modified it and put it in your own terms before expressing it to me, none of this reframing and modification—i.e. in biological terms, none of these acquired characteristics—would be present in the version you expressed to me. If ideas evolved through natural selection, I would get it in exactly the form your friend expressed it to you. This is clearly not what happens.[5]

Simonton claims, "Gabora's assertion [that Darwinism is incompatible with transmission of acquired traits] is historically incorrect…. [Darwin's] own version of the theory… was his theory of pangenesis, a theory that proposed a mechanism for the inheritance of acquired characteristics" (p. 388). The reality is that Darwin's all-but-forgotten notion of pangenesis is incompatible with the theory that made him famous, his theory of natural selection. Indeed, Darwin conceived of pangenesis in a drastic attempt to explain such oddities as limb regeneration, hybridization, and atavism, which he was extremely hard-pressed to make sense of without knowledge of modern genetics. However, it was clear to him that to the extent that pangenesis involves the transmission of acquired characteristics, adaptive change cannot be explained by natural selection, *i.e.* by what is now referred to as a Darwinian mechanism of change. That is, he was aware that change *across generations* in the frequencies of *heritable variations* in a population due to selection would easily be drowned out by a mechanism for the transmission of *acquired traits* operating *within a single generation*.

Simonton (2007b) claims "[Darwin's] chief departure from Lamarck was that he removed volition as a causal agent that produced the variation… Darwin's explicit goal was to provide a naturalistic account for the origin of species that was not contingent on special creation. That is the reason why his theory provoked such controversy (p. 388)." But there is nothing in Darwin's writings to indicate that he had any such agenda; his goal was simply to determine *how evolution works*. If the facts had led him to conclusions that were more consistent with "special creation" than any other theoretical framework, I imagine he would have accepted creationism as an explanatory framework. (That certainly would have made his job easier.) His repudiation of creationism was not merely symptomatic of a rebellious spirit; it reflects a devotion to the scientific enterprise and the facts before him. The issue as far as science is concerned is not why his theory provoked controversy amongst the religious, but why it *solved a scientific problem*. The question of whether biological change is driven by a creator may be important to Simonton, and to religious fanatics both now and in Darwin's day, but in scientific discussions it rarely get mentioned. To a scientist, the chief distinction between Darwinian and Lamarckian explanations is that in Lamarckian evolution acquired change is transmitted whereas in Darwinian evolution it is not.

## Darwinian, Neo-Darwinian, or Not Darwinian At All?

### Darwinian but Not Neo-Darwinian

A strategy used to salvage BVSR is to argue that creative thought is Darwinian, albeit not neo-Darwinian. In his reply to my commentary, Simonton (2007b) writes:

> [T]he new evolutionary synthesis could call itself neo-Darwinian, even though it repudiated the Lamarckian components in Darwin's thought. Hence, to condemn the BVSR theory of creativity as non-Darwinian is to commit a twofold error. First, Gabora introduces a double standard by permitting its usage for one phenomenon but not for another, when both uses are not completely faithful to Darwin's own formulation. Second, Gabora introduces an anachronistic criterion when she dismisses the BVSR theory of

> creativity because it does not comply with neo-Darwinism—a system that did not even exist in Darwin's day. (p. 389)

But there is no double standard here. Darwin considered the problem of evolution from many angles, and considered many possibilities, including Lamarckian ones, but these were not part of his theory of natural selection, and are in fact incompatible with it. The theoretical developments (e.g. population genetics) and empirical findings (e.g. discovery of DNA) associated with the neo-Darwinian synthesis did not in any way violate Darwin's theory of natural selection as an explanatory framework. Quite the contrary; they filled in important gaps. The synthesis did not change anyone's opinion about whether biological life evolves through a Darwinian process. The creative process, however, entails aspects that violate the applicability of natural selection as an explanatory framework, such as the transmission of acquired traits. Simonton argues that lack of inheritance of acquired traits is part of neo-Darwinian theory, but not part of Darwin's original theory (Simonton, 2007b). This is false. Darwin did not know *why* acquired traits are not inherited, but he observed that they are not, and that observation was an integral part of the foundation upon which his theory was built. It does not take knowledge of molecular genetics to observe that if an animal loses an ear, in a fight for example, its offspring are not born one-eared; this acquired trait is not transmitted. Darwin knew that to the extent that the distribution of variants in a population is due to transmission of acquired traits it is *not* due to natural selection. Natural selection, an *intra-generational* mechanism, would be quickly overwhelmed by a mechanism such as transmission of acquired traits operating *within* a generation.

In short, Darwin was fully aware that to the extent that acquired traits are inherited, natural selection cannot explain observed change. It is certainly not the case that creative thought must operate exactly like the evolution of biological species to evolve through natural selection. A process could violate some of these more detailed population genetics models and still be Darwinian, but it could not violate the basic principles of population genetics and still be Darwinian. Creative thought does not just violate the conditions for a neo-Darwinian explanation, it violates the conditions for even a Darwinian explanation, as Darwinism is standardly construed.

**Even Biological Evolution Was Not Originally Darwinian**

Simonton refers to my assertion that 'even biological evolution was not originally Darwinian as "extremely deceptive". He writes "This statement is based on her conception of what constitutes a Darwinian process, and her conception is plainly founded in what has been variously styled the modern evolutionary synthesis,… or simply neo-Darwinism" (p. 389). However, my conception of a Darwinian process is the standard conception universally accepted by biologists. Moreover, the findings I pointed to in this section of my commentary have nothing to do with the modern evolutionary synthesis. They show that the very earliest forms of life evolved through a process that *defies* the tenets of the modern synthesis. To the references originally cited as support (Bollobas, 2001; Bollobas & Rasmussen, 1989; Kauffman, 1993; Morowitz, 1992; Wächtershäuser, 1992; Weber, 2000; Williams & Frausto da Silva, 1999, 2003; Gabora, 2006; Vetsigian et al., 2006) others could be added. Some of the most intriguing include the work of National Medal of Science recipient Carl Woese and his colleagues. As Woese (2004) puts it: "Our experience with variation and selection in the modern context do not begin to prepare us for understanding what happened when cellular evolution was in its very early, rough-and-tumble phase(s) of spewing forth novelty" (p. 183). Vetsigian et al. (2006) write: "the evolutionary dynamic that gave rise to translation [the process by which the proteins that make up a body are constructed through the decoding of DNA] is undoubtedly non-Darwinian" (p. 10696). Woese and his colleagues show that early life underwent a transition from horizontal evolution through communal exchange, a process that is fundamentally cooperative, to vertical evolution through natural selection by way of the genetic code, a process that is fundamentally competitive. This transition is referred to as the *Darwinian threshold* (Woese, 2002) or *Darwinian transition* (Vetsigian et al., 2006). Kalin Vetsigian (pers. comm.) estimates that the period between when life first arose and the time of the Darwinian threshold spanned several hundred million years. Thus adaptive change is clearly *possible* through means other than selection.

**A Non-Darwinian BVSR**

A recent paper by Simonton suggests that after decades of claiming that creativity is Darwinian he may be backing down on this front, though continuing to develop a non-Darwinian version of BVSR (Simonton, 2010). However, a non-Darwinian BVSR is still a selectionist framework, and this kind of framework was developed to solve the paradox of how change accumulates despite being discarded at the end of each generation (Gabora, 2005). For example, you don't inherit your mother's tattoo; when your mother dies, your lineage loses this and all other traits she acquired during her lifetime. This paradox does not exist with respect to creativity.

Moreover, devoid of its Darwinian skeleton, it is not clear what BVSR can offer. The strength of BVSR was its connection to Darwinism, a conceptual framework with considerable explanatory power. Had it been possible to root creative thought in a Darwinian framework, we might arguably have achieved the first step toward a unification of the psychological sciences comparable with Darwin's unification of the life sciences. In the end, a non-Darwinian BVSR amounts to the assertion that creativity takes place in two stages, the first of which blindly generates ideas, and the second of which selects from amongst them. Of course, it is well known that this is not strictly the case; in practice, idea generation and exploration are strongly interleaved and intertwined. One could depart further from the Darwinian framework and modify the variation phase to incorporate certain biases, and modify the selection phase to include elaboration (as Simonton has). But then it is not clear what BVSR can provide beyond more conventional, well-known two-stage theories such as the "geneplore" (generate + explore) model (Fink, Ward, & Smith, 1992).

## Mis-representation of Alternative Theories

Simonton has a tendency to present alternative theories in a distorted form that bears little resemblance to the source. I will limit discussion of this to his version of my own theory of creativity, tentatively referred to as honing theory, from which the above alternative explanations for phenomena interpreted as support for BVSR were derived (Gabora, 2005, 2007, under revision). Simonton (2007b) writes:

> Let me begin by affirming that the honing of ideas does have a part to play in problem solving. On the one hand, honing does adequately describe what occurs in relatively

> routine problems that can be solved via well-established algorithms. Through a series of iterations that yield ever closer approximations, the individual converges on a solution. On the other hand, honing often proves useful in the later stages of creativity after the solution itself has been discovered. ... Gabora is right: Darwinian theory cannot explain either of these two honing processes. Even so, it doesn't have to. Neither of these activities would be considered highly creative by most researchers in the field. (p. 392)

Nowhere in any discussion of honing is there mention of "what occurs in relatively routine problems that can be solved via well-established algorithms" or "a series of iterations that yield ever closer approximations [by which] the individual converges on a solution" (p. 392). Nowhere in discussions of honing are the examples of it remotely similar to the scenario Simonton comes up with. Honing, as I use the term, refers to the simultaneous transformation of both (1) one's perspective on a creative task, and (2) one's internal model of the world, or *worldview,* in response to a problem, gap, sense of incompletion, or need to express oneself (Gabora, under revision). The process is driven by the self-organizing, self-mending tendencies of a worldview, and it need not occur in two stages; it can occur in one stage, or many. Through honing, the conception of the task typically shifts through interaction with real or imagined contexts, from *ill-defined*, in which some properties (primitive elements or components) are not specified, to *well-defined*, in which all key properties *are* specified.

A honing interpretation of the situation involving Picasso's sketch of the mother carrying the dead child is that the conception of the painting is in a dynamically unfolding state of potentiality that changes through observation of how a specific realization of a particular element of a painting, in the form of a sketch, affects the conception of that element, and of how that in turn affects the conception of the painting as a whole.

Not only are other theories misrepresented, but Simonton (2007b) distorts the degree to which others support BVSR. For example, he writes:

> "Gabora proclaims that even if BVSR might account for recombinations of already present ideas, it could not possibly account for innovative creations of the highest order … although I am pleased that Gabora is willing to acknowledge the place of BVSR processes in more mundane forms of creativity…" (p. 391).

This is misleading; I have never said anything to indicate that BVSR might account for recombinations of already present ideas, nor that it can account for mundane forms of creativity.

## The Status of BVSR as an Explanatory Framework for Creativity

It would be satisfying if a selectionist framework could explain as much about the process by which ideas take form in our minds as it does about the evolution of biological form. 'Blind variation selective retention', or BVSR, is an ambitious effort to achieve this. We examined the theoretical rationale for BVSR, and putative support based on analysis of a particular creative task, Picasso's painting of Guernica. Unfortunately, serious problems with BVSR come to light when we examine how this analysis has been interpreted.

Simonton has embellished BVSR to encompass expertise, remote associates, and the elaboration of creative ideas. Unfortunately, these embellishments, which have successfully enabled BVSR to match existing data on creativity, are in several respects incompatible with selectionist as an explanatory framework. Simonton claims that findings of false starts, backtracking, and so forth, in creative undertakings support the thesis that ideas are generated through trial and error, in large numbers, some of which are selected and many of which are not, and that one variant can be quite different from another. However, it is not intra-generational variation but change from one generation to the next that selectionism provides an explanation for, and BVSR has not been developed to the extent that it incorporates a means by which traits are passed on from one generation to the next, *i.e.,* a mechanism of inheritance. Darwinian theory does not even predict that data obtained from a particular creative task or analyses of lifetime creativity should exhibit nonmonotonicity. Darwin's theory of natural selection was put forth to explain why organisms become more adapted to their environment over time, so there is no reason that a lack of improvement in the quality or quantity of ideas across the lifespan *support* a Darwinian theory.

In loosening the original claim that variation is generated by chance, 'quasi-random' processes, Simonton has made BVSR more consistent with what we know about creativity, but he has inadvertently made it incompatible with the basic tenets of Darwinism. Expertise, logic, and activation of remote associates bias the generation of idea away from random. This renders BVSR inapplicable as an explanatory framework, because what is causing change is the *nature* of those biases, *not* selection: population-level shifts in the distribution of variants with heritable differences over generations due to competitive exclusion.

We have also seen that there are serious problems with the 'selective retention' component of BVSR. BVSR has been made more consistent with what we know about creativity through incorporation of the conscious elaboration of blindly generated variants into finished products. However, if these elaborations are shared with others this entails transmission of acquired traits, which again renders a selectionist explanation inapplicable. Selectionism is potentially applicable only to the extent that change accrued during a lifetime is obliterated at the end of each generation, because if change acquired over a lifetime is transmitted to offspring, this drowns out the slower population-level mechanism of change identified by Darwin. Moreover, the generation of one idea affects the conception of the task, and thus the criteria by which the next is judged. Therefore, successively generated ideas cannot be treated as members of a generation, and selected amongst. Thus not only is the notion of blind variation is problematic with respect to the creative process but so is the notion of selective retention.

Although BVSR does not provide an appropriate explanatory framework for creativity, it has made an important contribution to the study of creativity by directing our attention to a number of subtle and important issues. It has helped establish that steps toward the completion of a creative product exhibits nonmonotonicity, as does lifetime creativity, and that ideas do not arise through purely logical, rational processes. Ideas do not evolve through a selectionist process, but they clearly *evolve,* in the general sense of exhibiting descent with adaptive modification. Comparing and contrasting the process by which humans generate novel ideas with the process by which nature generates novel organic forms has provided valuable insights with which to move forward in our understanding of creative masterpieces such as Picasso's Guernica.

## Acknowledgements

I would like to thank Stefan Leijnen for comments on the manuscript. This work was funded in part by grants to the author from the *Social Sciences and Humanities Research Council of Canada* and the and the Concerted Research Program of the *Flemish Government of Belgium*.

## Notes

[1] The term 'quasi-random will be avoided here because—like 'quasi-infinite' or 'quasi-bottomless'—it is not clear what it means (although I can think of a meaning for 'quasi-topless'). The word pseudo-random, on the other hand, has been clearly defined. It refers to a sequence that is not truly random because it was generated by a deterministic process (for example, by a computer), but that exhibits statistical randomness and therefore cannot be distinguished by humans from random. (Because it is not genuinely random it can be used repeatedly to produce the exact same sequence of numbers.) If the generation of ideas were pseudo-random that would be sufficient, because the critical factor is whether it can be approximated by a random distribution. To the extent that the distribution of traits or characteristics deviates from a random distribution, natural selection gives a distorted model. Change over time is attributable to the source of this deviation, not to natural selection. However, the generation of ideas is by no means pseudo-random; it is biased by associative memory, drives, the present situation, and so forth. Therefore it cannot be approximated by a random distribution, and change over time cannot be modeled by natural selection.

[2] More precisely, it *is* possible for the theory of natural selection to be applicable even if the underlying process is not random. However, in that case, although not genuinely random, the process must be approximated by a random distribution. Biological variation is not genuinely random but the assumption of randomness generally holds well enough to serve as a useful approximation. (Actually, in some biological situations, such as assortative mating, the assumption of randomness does not hold, and in such cases natural selection is not an appropriate model.)

[3] Actually, some biological situations, such as assortative mating, cannot be accurately approximated by a random distribution, and to the extent that this is the case natural selection gives a distorted model.

[4] Nevertheless there are logistic reasons why change at the individual level *cannot* be highly nonmonotonic. One has to do with epistasis, wherein what allele is best at one locus depends on what allele is present at another locus. If the mutation rate is too high, evolution grinds to a halt because change at one locus affects not just that locus but the stable arrangement it has found with other loci to which it is epistatically linked. The bottom line is that nonmonotonicity in the creative process does not indicate that creativity is Darwinian.

[5] Note that mapping the idea to the genes is no less problematic than mapping the idea to the individual organism. This can be illustrated with a simple example. Say we genetically modify the iris of your eye such that it produces different pigment and your eye looks brown rather than blue. The probability that your offspring have the genes for blue eyes will still be as high as it was before this procedure. Blue lives on in this lineage. But consider the situation in which, in the course of painting a painting, you paint over a blue blob with brown paint. There is no mechanism in place that causes the painting to revert back to having the blue blob in it; i.e. that prohibits the brown paint from being present in future iterations of the painting, or from future paintings done in that style. Unless you actively scrape it off, the brown paint will continue to hide the blue blob. Blue is lost from the lineage. The reason for this is that the underlying mechanism of change in the two situations is very different (for further analysis, see Gabora, 2004, 2008).